\documentclass[pra,aps,twocolumn,superscriptaddress,longbibliography]{revtex4-2}

\usepackage{amsmath, graphicx, color, times, float, amssymb,xcolor}
\usepackage[colorlinks=true,citecolor=blue,linkcolor=blue,urlcolor=blue]{hyperref}
\usepackage{url}

\begin{document}

\title{Tilt-Induced Localization in Interacting Bose-Einstein Condensates for Quantum Sensing}

\author{Argha Debnath}
\affiliation{Harish-Chandra Research Institute, A CI of Homi Bhabha National Institute,
Chhatnag Road, Jhunsi, Allahabad 211 019, India}
\author{Mariusz Gajda}
\affiliation{Institute of Physics, Polish Academy of Sciences,
Aleja Lotnikow 32/46, PL-02668 Warsaw, Poland}
\author{Debraj Rakshit}
\affiliation{Harish-Chandra Research Institute, A CI of Homi Bhabha National Institute,
Chhatnag Road, Jhunsi, Allahabad 211 019, India}

\begin{abstract}
 We investigate localization transitions in interacting Bose-Einstein condensates (BECs) confined in tilted optical lattices, focusing on both the continuum limit accessed via shallow lattice depths and the tight-binding limit realized in the deep lattice regime. Utilizing the Gross-Pitaevskii equation (GPE) and the many-body Bose-Hubbard model, we analyze the scaling behavior of localization indicators, such as the root mean square width and fidelity susceptibility, as a function of the applied tilt. Our results reveal clear signatures of a localization-delocalization transition driven by the linear potential, with scaling properties that characterize criticality even in the presence of interactions within the GPE description. Despite the single-mode nature of the condensate wavefunction, we demonstrate that it can effectively probe quantum criticality. Building on this, we propose the use of interacting BECs in tilted lattices as a platform for quantum critical sensing, where the condensate wavefunction serves both as a sensitive probe of localization and a practical resource for quantum-enhanced metrology. This approach opens new avenues for precision gradient sensing based on localization phenomena in bosonic systems.
\end{abstract}

\maketitle

\section{\label{sec:intro}Introduction}

Localization, first identified by Anderson as the absence of diffusion due to disorder, has since emerged as a unifying concept in understanding how interference, interactions, and external potentials can inhibit transport in quantum systems \cite{anderson1958absence}. There have been several efforts to experimentally realize this exotic quantum phenomenon in various physical platforms. Ultracold atomic gases in optical lattices (OL) represent one of the most mature and versatile platforms in quantum simulation, offering unprecedented control over interactions, dimensionality, external potentials and disorder in engineered many-body systems \cite{lewenstein2007ultracold, lewenstein2012ultracold}. In particular, Bose-Einstein condensates (BECs) confined in OL offer a tunable setting in which the transition between extended and localized phases can be probed with exquisite control. BECs in OL have achieved groundbreaking success in demonstrating the phenomenon of localization. Experimental demonstrations of localization in BECs have spanned multiple systems that include noninteracting 
$^{87}$Rb condensates localized in 1D speckle potentials \cite{billy2008direct}, quasiperiodic localization of $^{39}$K
condensates in bichromatic lattices \cite{roati2008anderson}, 2D localization of $^{87}$Rb condensates in point-like disorder \cite{white2020observation}. Parallely, a plethora of theoretical studies have been conducted on the localization of the condensate trapped in the quasiperiodic potential
and random speckle potential \cite{adhikari2009localization,muruganandam2010localization,cheng2010symmetry,cheng2010spatially,cheng2010spatially,cheng2011localization,cheng2011matter,cheng2014localization,li2016localization,deissler2010delocalization,cheng2010matter,cardoso2012anderson,xi2015localization,zhang2022anderson}. Further numerical and experimental  studies have investigated the role of interactions in the BECs and have demonstrated that condensate can weaken localization, giving rise to a delocalized regime \cite{pikovsky2008destruction,kopidakis2008absence,lucioni2011observation,sarkar2023quench}.

Another complementary route to localization in the absence of disorder is through the application of a quasiperiodic \cite{aubry1980analyticity,iyer2013many,michal2014delocalization,modak2021many} or linear (tilted) potential, which breaks lattice translation symmetry and induces  localization.  In particular, localization occurs in the vanishing limit of the amplitude of the linear potential in the thermodynamic limit in the non-interacting systems. The studies on examining Stark localization have, primarily, been conducted on the tight-binding models. The research works include both single-particle Stark localization \cite{fukuyama1973tightly, holthaus1995random, kolovsky2008interplay, kolovsky2013wannier} and Stark many-body localization (MBL) \cite{ van2019bloch, schulz2019stark, wu2019bath, bhakuni2020drive, taylor2020experimental,morong2021observation,wei2022static}. Universality in the localization transition is characterized by critical exponents associated with system-dependent parameters, such as localization length. The single-particle works on the Stark localization include determining the critical exponents for comprehending the critical phenomena and the nature of the phase transitions within the tight-binding approximation \cite{Sachdev_2011, vojta2003quantum}. Whether disorder-free many-body systems can exhibit true localization remains a central question in the study of Stark MBL \cite{wannier1960wave}. Such studies have been corroborated with other interesting results, e.g., the emergence of an extensive set of quasilocal conserved quantities, called (quasi) local integrals of motion, in the Stark MBL \cite{bertoni2024local}.

Apart from the obvious phenomenological interest,  localization transitions have been recently advocated for applications in quantum metrology \cite{giovannetti2006quantum,degen2017quantum,oszmaniec2016random}. They belong to a larger class of quantum sensing devices called quantum critical sensors \cite{zhou2020quantum,montenegro2024review,mukhopadhyay2024current}. These efficient sensing tools exploit the vulnerability of a quantum many-body state near the transition point against a small shift of the parameter. The localization-delocalization transition has been demonstrated as a useful resource for quantum-enhanced parameter sensing in a number of settings that include localization induced solely either by quasi-periodic potential or Stark potential or by Stark potential with an additional quasi-periodic potential, in the fermionic tight-binding models and quantum spin chains \cite{modugno2009exponential,roy2020trends,yao2021many,he2023stark,liang2024quantum,sahoo2025stark,sahoo2024enhanced}. In particular, Stark systems can be used as a probe for the precise measurement of weak gradient fields and hence a promising platform for advancing quantum metrology, which represents a key frontier in the modern landscape of quantum technology.

In quantum metrology, quantum estimation theory provides a limit on how precisely an unknown parameter can be measured. Cram{\'e}r-Rao bound \cite{cramer1999mathematical,braunstein1994statistical} dictates the ultimate limit of precision in parameter estimation: ${\cal{E}}_Q \ge \left(M F_Q\right)^{-1}$, where ${\cal E}_Q$ is the uncertainty in the estimation of an unknown parameter, $F_Q$ is the quantum Fisher information (QFI) and $M$ is the number of repetitions in the measurement.  For small parameter shift QFI simply turns out to be scaled fidelity susceptibility \cite{rams2018limits}. The sensing performance is determined by the finite size scaling of the QFI or fidelity susceptibility, $F_Q \sim L^{\beta}$. $\beta = 1$ is  known as the standard quantum limit (SQL), which is the best that $L$ independent qubits can achieve \cite{giovannetti2004quantum,giovannetti2006quantum}, and a quantum-enhanced sensing implies $\beta > 1$. The so-called Heisenberg limit corresponds to $\beta = 2$ 
\cite{giovannetti2011advances,mondal2024multicritical}. Many-body probes that can reach the Heisenberg limit have been previously reported in different scenarios \cite{wei2019fidelity,desaules2021proposal,dooley2021robust,dooley2023entanglement,mishra2021driving,sarkar2022free,sahoo2024localization}. Recently, different aspects of quantum metrology have been explored in different context \cite{yang2024quantum,baak2024self,zhou2024limits,chen2024quantum,bhattacharyya2024even,abiuso2025fundamental,agarwal2025quantum}.

In this work, we investigate a degenerate Bose gas confined in a tilted OL, exploring both the tight-binding regime and the continuum-like regime accessible by tuning the lattice depth. BECs in optical lattices offer notable advantages over their fermionic counterparts. Apart from simpler cooling techniques, the long-range phase coherence inherent to BECs manifests directly in interference patterns, making them especially well-suited for studying phenomena such as superfluidity and localization. Our study is motivated by two central objectives. The first is to probe the quantum critical behavior associated with tilt-induced localization in interacting BECs. We approach this by analyzing scaling properties of physical observables such as the root mean square (RMS) width and fidelity susceptibility. These are computed using the Gross-Pitaevskii equation (GPE) for shallow lattices and through exact numerical simulations of the many-body Bose-Hubbard model in the tight-binding limit, realized at deep lattice depths. The second objective is to propose that tilt-induced localization transitions in BECs can serve as a powerful resource for quantum-enhanced sensing \cite{he2023stark}, particularly for detecting weak to intermediate gradient fields with high precision.

The structure of this paper is organized as follows. In Sec.~\ref{sec:model}, we present the governing equations to characterize the localization of ground states. Specifically, we analyze change in rms width with tilted OL potential strength and perform scaling analysis with system size in Sub sec.~\ref{subsec:nointer}. Furthermore, in Sub sec.~\ref{subsec:gpe fisher}, we discuss how the GPE based model can be borrowed for the development of advanced quantum sensors, particularly for the precise estimation of the Stark weak field amplitude. In sec.~\ref{sec:bh fisher} we performed similar analysis in  the tight-binding regime with the help of Bose-Hubbard model. Finally, we present our conclusions in Sec.~\ref{sec:conc}.
\begin{figure}[t!]
\includegraphics[scale=0.33]{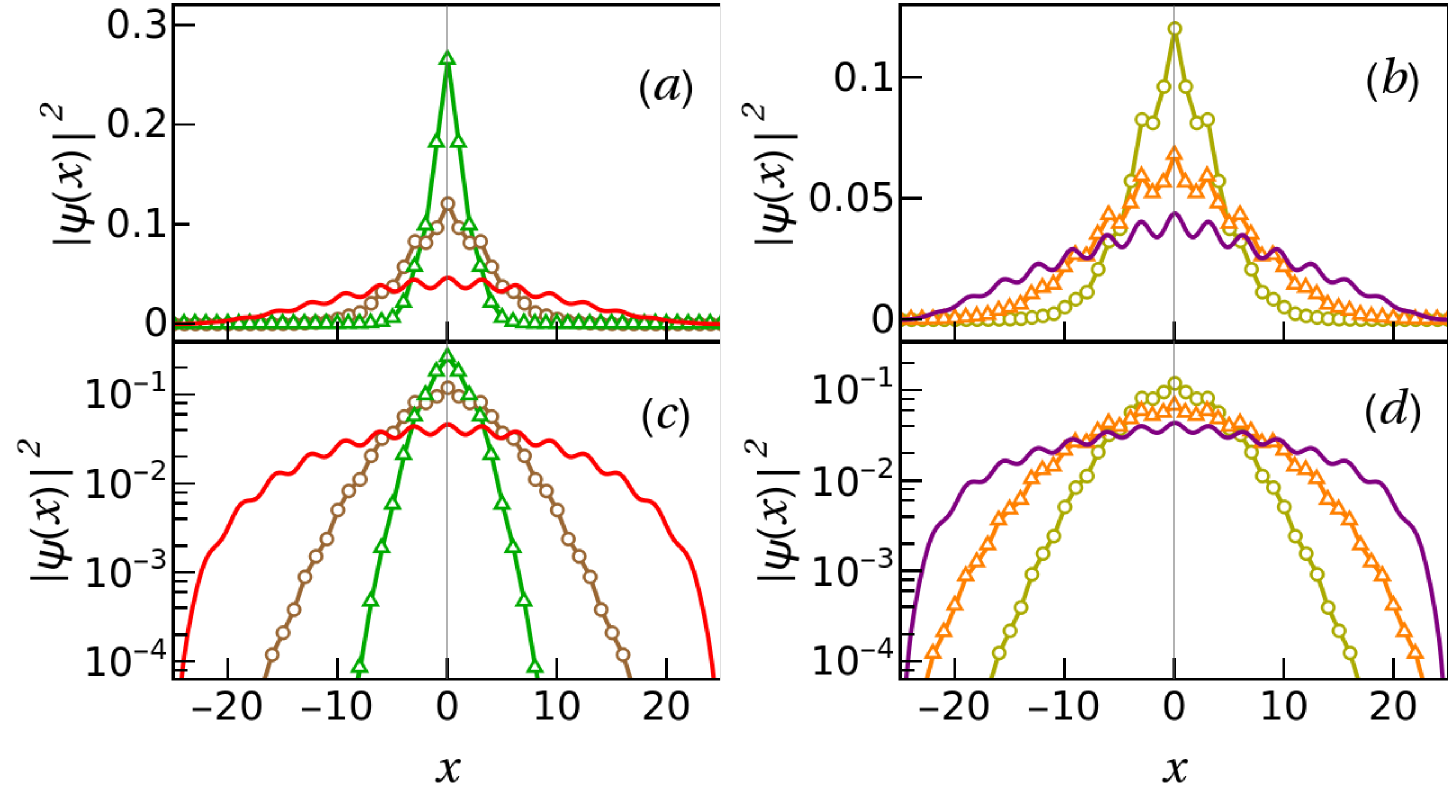}
\caption{Variation of density with three different relative Stark field strengths, $\tilde{V}=0.0002$ (red solid line), $\tilde{V}=0.02$ (brown circle), $\tilde{V}=0.2$ (green triangle up) for $g=0$. 
In (b) we showed variation of density for $\tilde{V}=0.02$ with $g$. Delocalization can be seen with the rise of interaction strength. Here, the plotted three different interaction strengths are, $g=0$ (yellow circle), $g=2$ (orange triangle up), $g=6$ (purple solid line). (c) and (d) are the log scale version of (a) and (b) respectively. We have taken $V=0.5$ and $L=50$. \label{fig:waveinterac}}
\end{figure}

\begin{figure*}[htbp]
\centering
\includegraphics[width=0.95\textwidth]{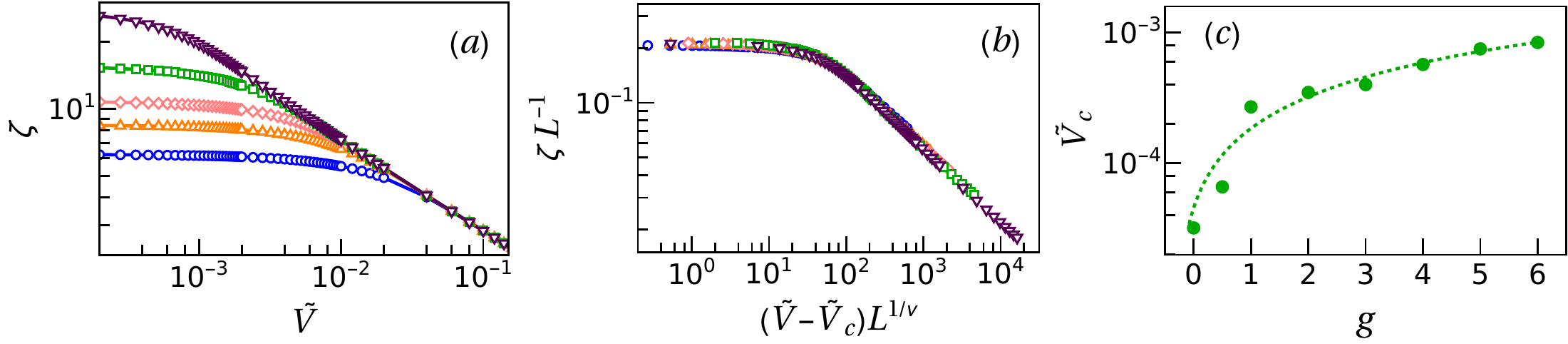}
\caption{For $g=1$ (a) presents RMS width,
$\zeta$ of the ground state against relative Stark field strength $\tilde{V}$ for different system sizes, $L=30$ (blue circle), $L=40$ (orange triangle up), $L=50$ (pink diamond), $L=70$ (green square) and $L=120$
(purple triangle down). (b) shows a collapse plot for the RMS width with the value of scaling exponent $\nu\sim 0.42$ and $\tilde{V}_c\sim 2.7\times 10^{-4}$. (c) Variation of critical Stark strength $\tilde{V}_c$ with interaction $g$. Green dashed line is added just to guide the eye of $\tilde{V}_c$'s incremental nature with increasing $g$. \label{fig:rms_all}}
\end{figure*}
\section{\label{sec:model} Tilt-Induced Effects in a Cold Bose Gas: Continuum Regime}
We consider the BEC loaded into a one-dimensional OL 
by the interference of counter-propagating two linearly polarized laser beams with wavelength $\lambda$ and associated wavevector $k$. The resulting periodic structure of the underlying trapping potential due to the OL is represented by $E \sin^2(kx)$, where the amplitude $E$ corresponds to the lattice depth. Such a lattice has been employed to investigate phenomena such as Bloch oscillations, Landau-Zener tunneling, mean-field effects, and lattice solitons \cite{efremidis2003lattice, cristiani2002experimental, mateo2011gap}. The OL is subjected to an additional $V$-shaped potential $f|x|$, where $f$ is a constant force. Hence, the BEC experiences the effect of a combined potential
\begin{equation}
V_{\text{ext}}(x) = E \sin^2(kx) + f |x|. \nonumber\\  
\end{equation}
When the optical lattice is superimposed on the V-shaped trap, the system can be viewed as two Stark-type tilted lattices with opposite slopes joined at a cusp at 
$x=0$.  Such a V-shaped potential is experimentally realizable using magnetic quadrupole trapping fields. In order to minimize the Zeeman energy, the atomic magnetic moment aligns along the external magnetic field, $\vec{\sigma} \| \vec{B}$, and the energy
is equal to $E=- \sigma |B(x,yz)|$. The magnetic moment of ultracold atoms follows adiabatically the local direction of the magnetic field; thus, a linear magnetic field gives the Zeeman energy equal to $E=-\sigma |x|$, thereby producing an effective potential of the form 
$f|x|$, where $f$ is determined by the magnetic-field gradient and the atomic magnetic moment.

Repulsive mean-field nature is introduced through the effective 1D interatomic interaction constant $g_{\text{int}}$ which depends on the transverse confinement via $g_{\text{int}}=\frac{4 \hbar^2 a_s}{m a_{\perp}^2}\left(1-C a_s/a_{\perp}\right)^{-1}$, where $m$ is the atomic mass, $a_s$ is the 3D scattering length, $a_{\perp}$ is the transverse confinement length, and $C \approx 1.4603$ \cite{olshanii1998atomic}. 
The total atom number is given by $N = \int dx |\tilde{\psi}(x,t)|^2$, where $\tilde{\psi}$ is a complex scalar field representing the interacting BEC order parameter. It is convenient to work with the rescaled field such that $\psi(x,t) = \sqrt{N} \tilde{\psi}(x,t)$ with $\int dx |\psi(x,t)|^2 = 1$. 
One obtains a non-linear GPE of the form 
\begin{eqnarray}
i \partial_t \psi=(-\partial^2_x/2+V \sin^2(x)+V_0|x|+g|\psi|^2)\psi,\label{eq:one}
\end{eqnarray}
which is presented in a dimensionless form by rescaling the associated length scale of the system as $x \to x k$, the time as $t \to 2E_k t/\hbar$, and $\psi$ as $\psi \to \psi/\sqrt{k} $, where $E_k = \hbar^2 k^2/2m$ is the recoil energy. Here  $V=E/(2E_k)$, $V_0=f/(2kE_k)$ and $g = N g_{\text{int}}/(2 E_k)$ are the normalized dimensionless parameters. Rather than considering interaction $g_{int}$ and atom number $N$ separately, it is advantageous to present the results by assigning $g \sim g_{int} N$ that can be easily used to simulate different experimental situations with different interactions and distinct bosonic atoms. Nevertheless, the effective nonlinearity $g$ can be experimentally varied over a broad range by either controlling the atom number or by tuning the scattering length via magnetic field near a Feshbach resonance~\cite{inouye1998observation}. The stationary solution of the  GPE (Eq.~\ref{eq:one}) is obtained numerically through the imaginary-time evolution technique. We consider the BEC to be confined in a finite OL of size $L$: $x \in [-L/2,L/2]$; and it is subjected to a closed boundary condition with $\psi(\pm L/2) = 0$. 

In the absence of an applied tilt (i.e., with  $V_0=0$) and in a weakly modulated OL, the GPE admits extended stationary states analogous to Bloch waves, which span the entire optical lattice and reflect its translational symmetry. In this study, we choose $V < 1$, corresponding to the height of the potential well barriers. This choice is motivated to keep the system away from the tight-binding limit, which is typically associated with $V > 10$ \cite{ cristiani2002experimental}. More specifically, we impose the kinetic energy constraint $\kappa^2 = 2 (V- \mu) < 1$ \cite{efremidis2003lattice}, where $\mu$ is the chemical potential, to ensure that the system operates within a continuum-like regime amenable to the treatments of the GPE. This setup enables an isolated investigation of how a linear potential (the tilt) affects localization–delocalization behavior in the presence of a shallow OL. Within this regime, the GPE is expected to capture the onset of localization driven by the applied tilt. As we demonstrate in the following section, increasing $V_0$ leads to progressive localization of the condensate. As the lattice depth increases and the tight-binding limit is approached, one can expect fragmentation of the BEC in the absence of the applied tilt. The applicability of the GPE is compromised in this regime, where many-body effects dominate and mean-field coherence breaks down. To explore this regime, we employ numerical simulations, particularly suitable for accurate estimation of the low-energy physics of the fractionally filled, interacting Bose-Hubbard model in the titled OL. This allows us to probe  many-body localization–delocalization transitions in finite systems, going beyond the mean-field description.

\subsection{\label{subsec:nointer}Localization-delocalization and scaling analysis}

To understand the nature of localization-delocalization transition in the finite systems and how they respond to various system parameters, we begin by examining delocalized states that arise at low values of the tilt strength $V$, and then fine-tune it to the larger values that correspond to the localized phase. We then study how repulsive interactions, parameterized by the coupling constant, $g$, influence these localized states. The inclusion of a repulsive nonlinearity in Eq.~\ref{eq:one} significantly suppresses the emergence of localization at values of $V$ where localization would occur in the absence of interactions. For example, a condensate of approximately $g \sim 6$ significantly inhibits Anderson localization in presence of a random potential~\cite{cheng2010matter,wang2000ground}. 

In order to perform a systematic analysis of localization we vary $\tilde{V}=V_0/V$ and the strength of nonlinearity $g$, keeping $V$ fixed at $V=0.5$. We monitor two specific situations. First, we vary $\tilde{V}$ for a fixed $g$. In finite systems, a delocalization-localization transition is expected for finite $\tilde{V}$. Second, we begin by initiating the system in the localized phase by setting $\tilde{V}$ at a finite value that is beyond the transition value, and then gradually increasing the non-linear repulsive interaction strength. This is supposed to weaken the localization effect. We illustrate these two situations in Fig.~\ref{fig:waveinterac}. In Figs.~\ref{fig:waveinterac}(a) and (c), we plot the results of spatial density distribution from numerical analysis for system size $L=50$ for the non-interacting case ($g=0$). It can be noticed that the  system initially has an extended nature of the condensate at a weak $\tilde{V}$ value. It gets more and more localized with increasing tilt strength. This is marked by a gradual enhancement of the density around the central region and the disappearance of the condensate density at the edges. We demonstrate the effects of enhanced interaction on a localized BEC in Figs.~\ref{fig:waveinterac}(b) and (d) for the same system size. Gradual spatial inflation of the condensate can be observed with increasing $g$ until it gets completely extended at a sufficiently high interaction strength.

In order to formally understand the nature of the  localization transition and associated critical properties in the thermodynamic limit, we compute the root-mean square (RMS) width, $\zeta$, which is defined as,
\begin{eqnarray}
\zeta^2=\int_{-\infty}^{\infty} (x-\langle x\rangle)^2|\psi(x,t)|^2 dx,\label{eq:four}
\end{eqnarray}
where, $\langle x \rangle=\int_{-\infty}^{\infty} x|\psi(x,t)|^2 dx$. We study the dependence of $\zeta$ for a finite interaction strength $(g=1)$ on the control parameter $\tilde{V}$ for different system sizes $L$ in Fig. \ref{fig:rms_all} (a). An initial flat region, implying an extended nature of the BEC, can be noticed in the finite-sized systems. After exceeding a certain threshold of $\tilde{V}$, say $\tilde{V}_T$, the effect of system size on  $\zeta$ almost lost. This marks system's entrance into the localized regime. The threshold value $\tilde{V}_T$ gradually decreases with increasing system-size. The RMS width follows the relation $\zeta\propto |\tilde{V}-\tilde{V}_c|^{-\nu}$, where the scaling exponent $\nu$ determines the rate of divergence of the RMS width in the thermodynamic limit near criticality and $\tilde{V}_c$ is the critical strength of the tilt corresponding to the localization in the thermodynamic limit, i.e., $\tilde{V}_c = \lim_{L \to \infty} \tilde{V}_T$. The scaling exponent, $\nu$, and $\tilde{V}_c$ can be obtained via two methods. A straightforward method is to perform a direct fitting analysis. Extraction of the scaling exponents via the well known technique of data collapse is another alternative method \cite{modak2021finite}. To determine scaling exponent of the RMS width via data collapse, we adopt the following
scaling ansatz,
\begin{eqnarray}
\zeta=&&L f_1\left((\tilde{V}-\tilde{V}_c)L^{1/\nu}\right),\label{eq:five}
\end{eqnarray}
where $f_1[.]$ is an arbitrary function and $\tilde{V}_c = \lim_{L \to \infty} \tilde{V}_T(L)$.  
As shown in Fig. \ref{fig:rms_all}(b), the curves collapse onto each other for $(\nu, \tilde{V}_c) \sim \left(0.42, 0.00027\right)$. Similar analysis for the non-interacting case gives $\nu\sim 0.33$. In the absence of interaction ($g=0$), the scaling exponent obtained in the continuum limit for a shallow lattice closely matches the known result for a single particle in a nearest-neighbor tight-binding lattice subject to a Stark potential. In the thermodynamic limit, the critical potential approaches  zero, as expected. However, this lies beyond the resolution limit of the GPE simulations due to numerical constraints. It is evident that the presence of interaction leads to an increase in the value of $\zeta$ for any given $\tilde{V}$. 
The effect of $g$ on $\tilde{V}_c$ is exhibited in Fig.~\ref{fig:rms_all}(c). The analysis suggests that, in the thermodynamic limit, interactions shift the localization threshold to higher values of the linear potential.

\subsection{\label{subsec:gpe fisher}Finite-size scaling of Quantum Fisher Information and Quantum Sensing}

If an unknown parameter $\tilde{V}$ is encoded in a probe state
$\psi(\tilde{V})$, the uncertainty of the parameter  near the $\tilde{V}_c$ is captured
by the fidelity susceptibility $\eta_Q$, which is defined as
\begin{equation}
    \eta_Q=-\lim_{\delta \tilde{V} \to 0} \frac{\partial^2 \mathcal{F}}{\partial (\delta \tilde{V})^2},    
\end{equation}
where
$\mathcal{F}(\tilde{V}, \delta\tilde{V})=\langle\psi(\tilde{V})|\psi(\tilde{V}+\delta\tilde{V})\rangle$ is the fidelity. The QFI is related
to susceptibility to fidelity as $F_Q=4\eta_Q$ \cite{rams2018limits}. Quantum states belonging to different quantum phases exhibit characteristically distinct properties. Many-body quantum-sensing protocols essentially exploit the sudden change in these properties near a quantum critical point, a feature that is effectively captured by the QFI.  Although within the classical field ($c$-number) approximation, the condensate wavefunction is a single-mode description; it effectively detects the localization transition, as shown previously. This motivates its use not only as a probe of criticality but also as a practical resource for designing quantum critical sensing device. It is well justified to utilize single-mode QFI for quantifying the sensitivity in the estimation of an unknown parameter.
\begin{figure}
\includegraphics[scale=0.35]{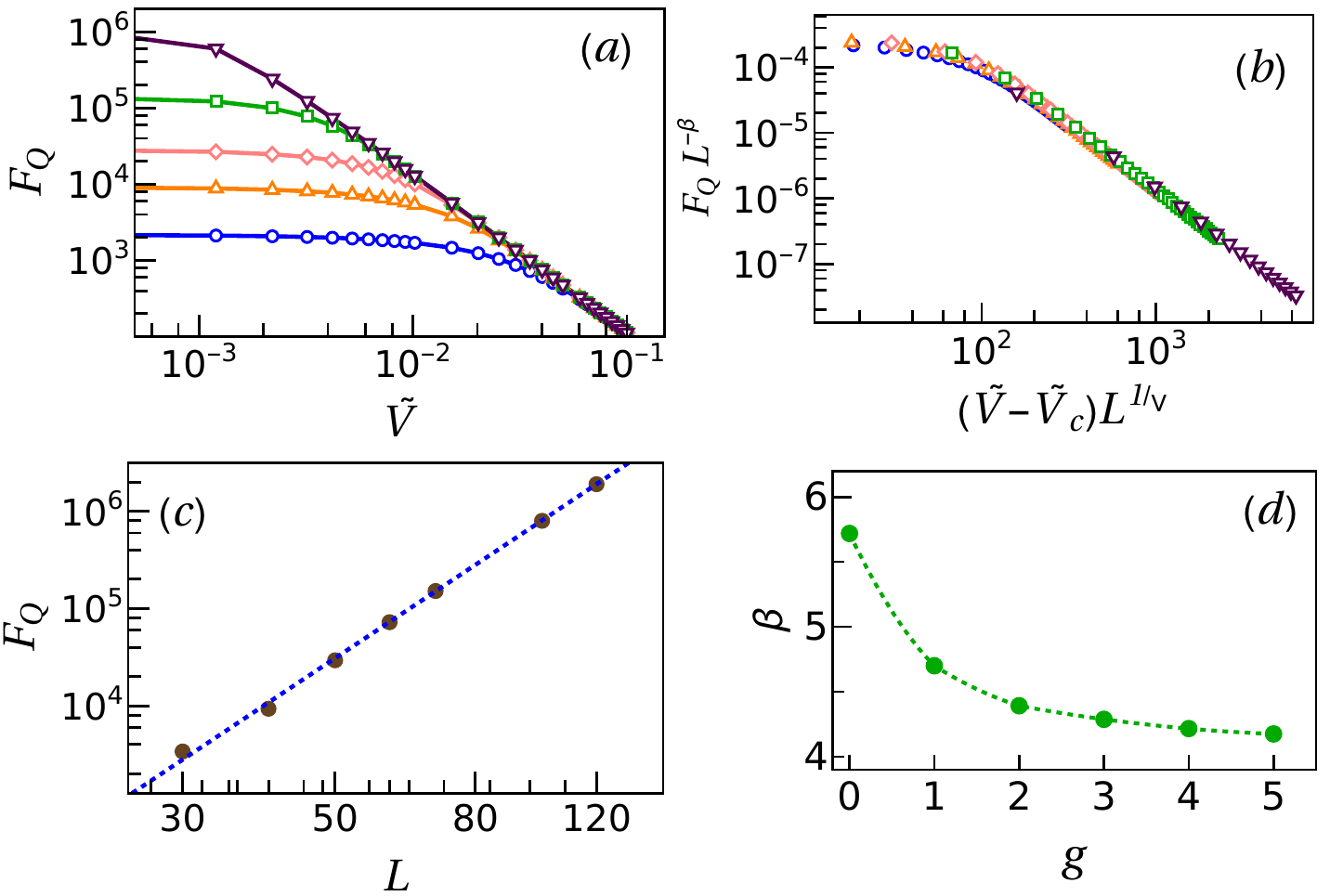}
\caption{For $g=1$ (a) presents The QFI, $F_Q$ versus relative Stark field strength $\tilde{V}$ for different system sizes, $L=30$ (blue circle), $L=40$ (orange triangle up), $L=50$ (pink diamond), $L=70$ (green square) and $L=120$
(purple triangle down). (b) shows a collapse plot for the $F_Q$ with $(\beta, \nu, \tilde{V}_c)\sim(4.7,0.42, 2\times 10^{-4})$. (c) QFI (brown dots) as a function of $L$ for a fixed $\tilde{V}$. The dashed
blue lines are fitting of the form $F_Q(\tilde{V}=10^{-5})\propto L^\beta$ with $\beta\sim4.73$. (d) we have plotted variation of $\beta$ with $g$.\label{fig:fisher_all}}
\end{figure}

We plot QFI, $F_Q$, as a function of $\tilde{V}$ for various $L$ and $g=1$ in Fig.~\ref{fig:fisher_all}(a). 
 The $F_Q$ remains nearly flat in the superfluid regime for $\tilde{V}<\tilde{V}_T$. Finite-size effects are evident in these initial plateaus of the QFI, representing the extended phase of the system. Following are the key features: First, by increasing $L$, the value of the QFI dramatically enhances. Second, the position of the maximum QFI value gradually shifts toward lower $\tilde{V}_T$ with increasing $L$. As previously shown, the act of $\tilde{V}_c\to 0$ in thermodynamic limit for non-interacting case is violated in presence of interaction. Third, in the localized regime, after a certain threshold $\tilde{V}>\tilde{V}_T$, the QFI becomes nearly size-independent in the localized phase.  We propose the following ansatz for extracting the associated scaling exponents via the data collapse,
\begin{eqnarray}
F_Q=&&L^{\beta}f_3\left((\tilde{V}-\tilde{V}_c)L^{1/\nu}\right),\label{eq:six}
\end{eqnarray}
where $f_3[.]$ is an arbitrary function. Fig.~\ref{fig:fisher_all}(b) represents the collapse plot of the QFI for $g=1$. The scaling exponents $\beta$ and $\nu$ turn out to be 4.7 and 0.42, respectively, whereas the value of $\tilde{V}_c$ is obtained as $V_c\sim 0.0002$. 

In the many-body context, system-size is a resource, and hence, the scaling of  $F_Q$, with system size, $L$, is of key interest. In the extended phase, the QFI is highly dependent on size $L$. To see how it scales with the probe size, in Fig.~\ref{fig:fisher_all}(c) we plot the QFI at extended phase for a fixed $\tilde{V}=10^{-5}$ as a function of $L$. The QFI is shown by brown dots, and
blue dashed lines are fitting functions $F_Q\propto L^\beta$ with
$\beta\sim4.73$. This shows super-Heisenberg scaling have also been proposed recently on the quantum many-body platform \cite{mondal2024multicritical,sahoo2024enhanced,sahoo2025stark,agarwal2025critical}. Fig.~\ref{fig:fisher_all}(d) displays the exponent as a function of $g$. The trend suggests that it, $\beta$, decreases with increasing strength of the nonlinear interaction $g$.
 Although we employ a symmetric linear (V-shaped) Stark potential that preserves spatial reflection symmetry, the sensing performance and scaling behavior derived in this work do not rely on the specific symmetry of the 
$|x|$ profile and remain robust in the presence of uniform linear gradients. In particular, we have verified that replacing the symmetric potential $V_0 |x|$ by a conventional linear tilt 
$V_0 x$ in Eq.~\ref{eq:one} leads to qualitatively identical conclusions (see Appendix for details).

\begin{figure}
\includegraphics[scale=0.34]{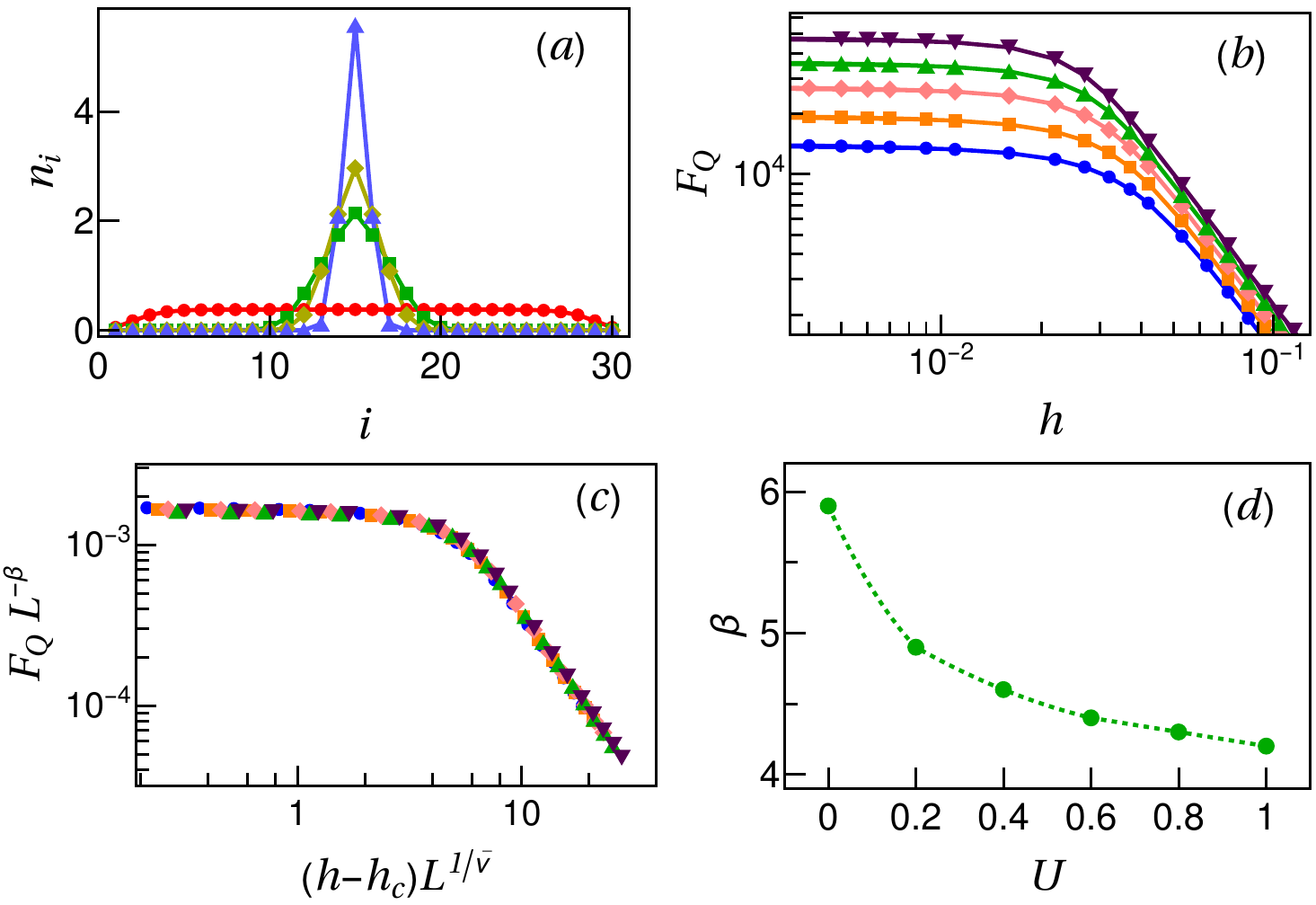}
\caption{(a) Density profiles $n_i$ for different Stark potentials $h=0$ (red filled circle), 0.5 (green filled square), 1 (yellow filled diamond), 4 (light blue filled uptriangle) for $L=30$. (b) presents the QFI, $F_Q$ versus Stark field
strength $h$ for different system sizes, $L=36$ (blue filled circle), $L=39$
(orange filled triangle up), $L=42$ (pink filled diamond), $L=45$ (green filled square) and $L=48$ (purple filled triangle down). (c) shows a collapse plot
for the $F_Q$ with the value of scaling exponents $(\beta, \bar{\nu}, h_c)\sim(4.3, 0.71, 3\times 10^{-3})$. (d) Variation of $\beta$ with $U$. Each $\beta$ is obtained through fitting of the form $F_Q(h=10^{-3})\propto L^\beta$. For (a), (b) and (c) interaction strength is taken as $U=t=1$.\label{fig:bosehuball_all}}
\end{figure}

\section{\label{sec:bh fisher}TILT-INDUCED EFFECTS IN A COLD BOSE GAS: Tight-Binding Limit} 
In the deep lattice regime, nearest-neighbor tunneling becomes increasingly suppressed, and tunneling to more distant sites is negligible up to exponentially small corrections. In the tight-binding limit within the lowest-band approximation, the system of ultra-cold bosons in the optical lattice is well approximated by the standard Bose-Hubbard model. The corresponding Hamiltonian, $\hat{H}_1$, is given by
\begin{eqnarray}
    \hat{H_1} = -t \sum_i \left(\hat{b}_i^{\dagger} \hat{b}_{i+1} + \text{h.c.}\right) 
    + \frac{U}{2} \sum_i \hat{n}_i (\hat{n}_i - 1),
\end{eqnarray}
where $t$ is the nearest-neighbor tunneling amplitude and $U$ denotes the  the onsite interaction strength.  The interaction parameter $U$ is given by an integral over the exact Wannier on-site wavefunction $w(x)$ of the lowest band and is given by $U = g \int dx |w(x)|^4$. The overlap between the nearest-neighbor wavefunctions dictates the tunneling rate. In the deep-lattice limit $E >> E_k$, the interaction strength increases with increasing $E$ due to the shrunken width of the Wannier functions, and, at the same time, the kinetic energy experiences an exponential decay. Hence, the strongly correlated regime $U/t \sim 1$ can be realized by tuning the depth of the optical lattice potential. In presence of an additional tilting potential, the Hamiltonian gets modified as 
\begin{equation}
\hat{H}=\hat{H}_1 + h \hat{H}_2,    
\end{equation}
where $h$ is a constant parameter and $\hat{H}_2$ corresponds to the tilting potential, the specific form that is considered to be
\begin{equation}
    \hat{H}_2 = \sum_i |(i-i_c)| \hat{b}_i^{\dagger} \hat{b}_i,
\end{equation}
where $i_c$ refers to the central site. Specifically, for a system with $L$ sites, we take $i_c=L/2 \left((L+1)/2\right)$ for even(odd) $L$.  Effects of tilt in Bose-Hubbard model has already been studied in different context \cite{taylor2020experimental,yao2020many,khemani2020localization}.

We consider fractionally filled latices subjected to repulsive interaction. In particular, we keep the filling fixed at 1/3 throughout this work. The fractionally filled system typically remains in the superfluid phase as it lacks the commensurability needed for the Mott insulating state. The correlated bosonic system at $h=0$ remains in a superfluid state (a gapless Tomonaga-Luttinger liquid).
In the single-particle case, all eigenstates of the non-interacting system are localized in the presence of the tilting potential in the limit of vanishing $h$, that is, $h \to 0$, in the thermodynamic limit, $L \to \infty$. We study the delocalization-localization crossover in finite-sized systems in the presence of interaction. In particular, our interest remains in the ground state behavior of the QFI, or fidelity-susceptibility, for monitoring the finite-size crossovers. We extract relevant scaling exponents that essentially characterize the low-energy universal behavior of the localization transition in the thermodynamic limit. As discussed before, the finite-size scaling of the QFI determines the usefulness of a quantum many-body system for estimating an unknown parameter. The results for the finite-size interacting system are obtained by performing density matrix renormalization group (DMRG) calculations via matrix product state (MPS) formalism. We consider an open boundary condition for maintaining the desired accuracy in the calculations. Below we present the results. We set $t=1$ for convenience.

In Fig.~\ref{fig:bosehuball_all}(a), we show the density profile of the system at 1/3 filling for a fixed system size $L=30$, interaction strength, $U = 1$, and varied strengths of the Stark potential $h$ covering a wide range. One can observe the gradual transition from the extended to the localized nature of the system as $h$ increases in magnitude. In order to investigate the delocalization-localization crossover in finite-size systems, we investigate QFI, which is also the scaled fidelity-susceptibility (see \ref{subsec:gpe fisher}). Fig.~\ref{fig:bosehuball_all}(b) illustrates QFI, $F_Q$, as a function of $h$ at $U=1$ for different system sizes. $F_Q$ is characterized by an initial flat region beyond which, that is, beyond a certain threshold of $h$, say $h=h_T$,  $F_Q$ exhibits decay with $h$ that marks the system's entrance into the localized phase. The initial flat region implies the extended nature of the many-body wavefunction below $h_T$. Near criticality, the QFI scales with system size as $F_Q \propto L^{-\beta}$. In order to extract the scaling exponent, we use following scaling ansatz
\begin{equation}
    F_Q = L^{\beta} f_4\left((h-h_c) L^{1/\bar{\nu}}\right),
\end{equation}
where $h_c = \lim_{L \to \infty} h_T(L)$ is the critical point and $f_4[.]$ is an arbitrary function. The validity of the scaling approach can be confirmed from the collapse plot depicted in Fig.~\ref{fig:bosehuball_all}(c). The scaling exponents are unsheathed from the data collapse, which suggests $\beta\sim4.3$, $\bar{\nu}\sim0.71$ and $h_c\sim0.003$ at $U=1$. The scaling exponent $\beta$ is extracted by employing a combination of the direct fitting and the data collapse techniques for various interaction strengths. $\beta$ turns out to be $\beta \sim 6$ for the non-interacting case, i.e., $U=0$. It gradually declines with increasing interaction strength. The trend is illustrated in Fig.~\ref{fig:bosehuball_all}(d). Given that $\beta>2$ for the considered cases in the tight-binding limit as well, the system achieves a super-Heisenberg scaling with the system size. Thus,  it is possible to design efficient quantum critical sensing device via a tilted bosonic lattice for estimating an unknown parameter, which is in this case a weak field parametrized by $h$.

\section{\label{sec:conc}Discussion}
We have investigated a trapped BEC in an OL under a linear (Stark) potential, focusing on the interplay between tilt-induced localization and finite-size effects. Using the GPE, we show that the system undergoes a crossover from an extended to a localized density profile as the tilt increases, with the RMS width serving as a sensitive diagnostic. This crossover sharpens with increasing system size and becomes approximately size-independent beyond a critical field strength. Repulsive interactions counteract localization but do
not eliminate the transition. Scaling analysis reveals that the critical behavior is governed by exponents close to those of
the non-interacting Stark-localized system. We further examine QFI as a metrological witness, demonstrating super-Heisenberg scaling in the extended regime. Complementary
Bose-Hubbard simulations in the tight-binding regime support this picture and indicate that the superfluid phase retains continuum-like sensitivity even at moderate interactions and
lattice depths. Although the applicability regimes of GPE and Bose–Hubbard are different, both of these descriptions are well suitable for describing the underlying physics of fully-fledged complex Schr{\"o}dinger equation, their domains of validity are dictated by microscopic details such as lattice depth and interaction strength; however, in the extended (superfluid) regime, the relevant length scales far exceed the lattice spacing. As a consequence, microscopic details, including short-range structure and precise lattice parameters, become irrelevant, and different microscopic models converge to the same low-energy, universal behavior. As a consequence, both the descriptions give similar behavior in terms of localization length in the extended regime. Importantly, it is precisely within this extended phase, and in particular in its proximity to the finite-size localization–delocalization crossover, that the metrological advantage and enhanced scaling of the QFI emerges. In contrast, once the system enters the localized regime, observables saturate, becoming essentially independent of system size, so no scaling advantage remains. The onset of the localized phase therefore serves to mark the boundary of the useful sensing window, while the quantum enhancement arises from the universal physics of the extended phase approaching localization.

Our findings underscore the utility of tilt-induced localization as a tunable probe of criticality in interacting bosonic systems. Despite its mean-field nature, the GPE captures key features of the transition, suggesting that localized modes in
this setting could serve as building blocks for quantum sensors or gradient field detectors. Beyond sensing, this work
connects to foundational questions in many-body localization without disorder and offers a fertile platform for exploring
nonequilibrium quantum phase transitions. Investigations of time-resolved response to tilt ramps, thermal effects or quasi-particle excitations, and generalization to multicomponent or spinor condensates would be natural extensions of this work.  Moreover, while the static scaling analysis presented here focuses on equilibrium signatures,  the dynamical signature of the Stark localization is expected to be manifested via strong suppression of transport and emergence of long-lived nonergodic dynamics \cite{yao2020many,morong2021observation}. In future it will be interesting to explore the possiblity of engineering dynamical quantum sensors \cite{sahoo2024localization} by exploiting these
unique dynamical features due to the Stark localization.

\appendix
\setcounter{figure}{0}
\renewcommand{\thefigure}{A\arabic{figure}}

\section{\label{sec:app}Analysis with Linear Tilt}
\begin{figure}[h!]
\centering
\includegraphics[scale=0.35]{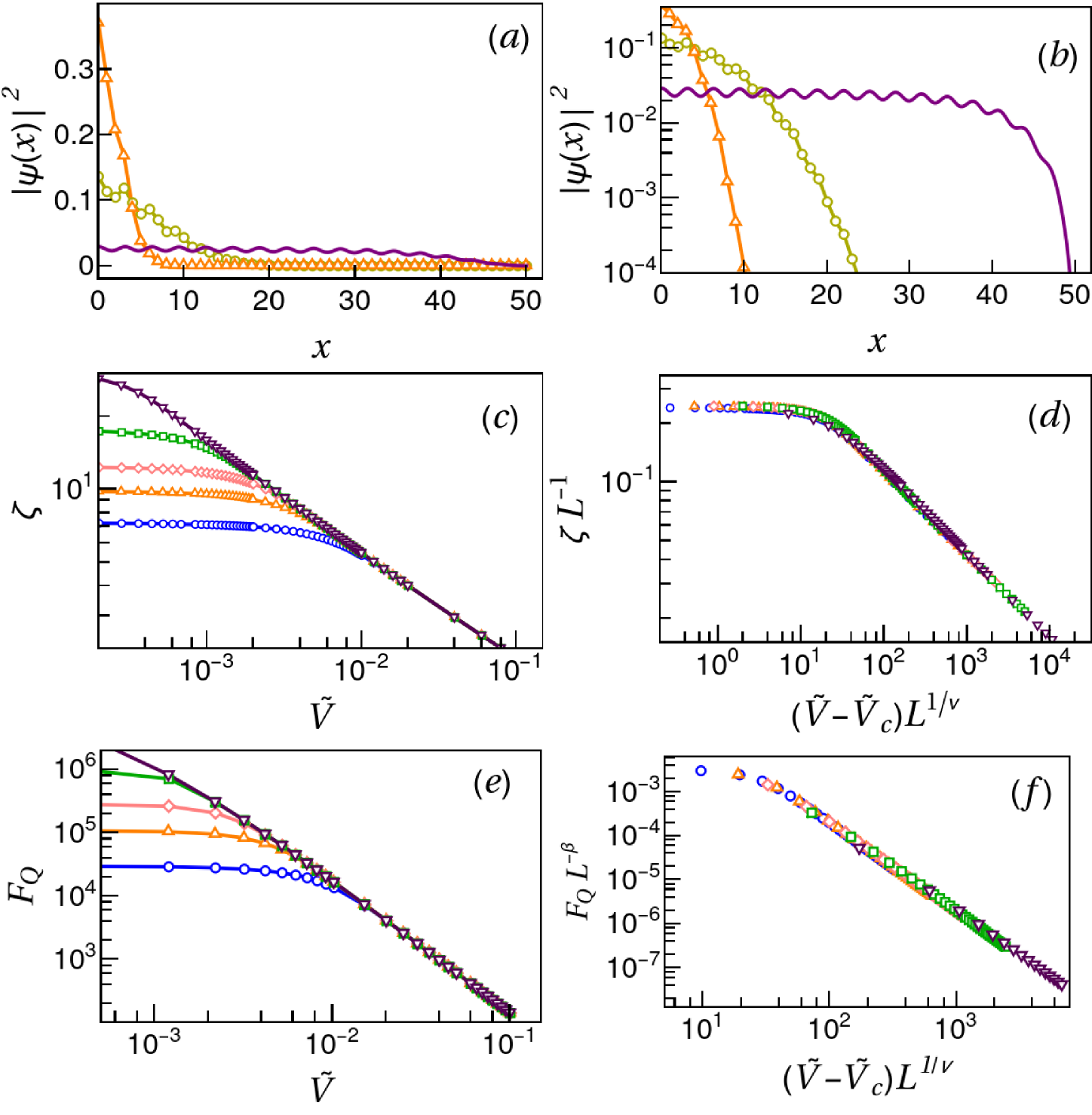}
\caption{ Figure depicts the results with linear tilt. Figure (a) displays the variation of density with three different relative Stark field strengths, $\tilde{V}=0.0002$ (purple solid line), $\tilde{V}=0.02$ (yellow circle), $\tilde{V}=0.2$ (orange triangle up). Log scale counterpart of (a) is plotted in (b). (c) presents RMS width, $\zeta$ of the ground state against relative Stark field strength $\tilde{V}$ for different system sizes, $L=30$ (blue circle), $L=40$ (orange triangle up), $L=50$ (pink diamond), $L=70$ (green square) and $L=120$ (purple triangle down). (d) shows a collapse plot for with the value of scaling exponent $\nu\sim 0.42$ and $\tilde{V}_c\sim 2.7\times 10^{-4}$. (e) presents The QFI, $F_Q$ versus relative Stark field strength $\tilde{V}$ for different system sizes. (d) and (f) show collapse plots for the RMS width and $F_Q$ with $(\beta, \nu, \tilde{V}_c)\sim(4.7,0.42, 2\times 10^{-4})$, respectively. }\label{fig:app_all}
\end{figure}

For completeness, we examine conventional uniform linear tilt, replacing the symmetric Stark potential $V_0 |x|$ considered in the main text by a linear gradient $V_0 x$. This allows a direct comparison between the V-shaped and uniformly tilted lattices. As we show below, the principal conclusions remain qualitatively unchanged, indicating that the observed effects do not rely on the reflection symmetry of the V-shaped potential but rather on the presence of a spatially varying Stark lattice itself.

We consider the condensate to be confined in a finite optical lattice of size $L$. In presence of the linear tilt, Eq.~\ref{eq:one} is modified as 
\begin{eqnarray}
i \partial_t \psi=(-\partial^2_x/2+V \sin^2(x)+V_0x+g|\psi|^2)\psi,\label{eq:app}
\end{eqnarray}
Here, we subject the condensate to boundary conditions: $\psi^\prime(x)|_{x=0}=0$ and $\psi(x)|_{x=L}=0$. We perform similar analysis with localization varying  $\tilde{V}=V_0/V$ keeping $V$ fixed at $V=0.5$.  We vary $\tilde{V}$ for a fixed $g=1$. In finite systems, a delocalization-localization transition is expected for finite $\tilde{V}$.  We illustrate the results in Fig.~\ref{fig:app_all}. In Figs.~\ref{fig:app_all}(a) and (b), we plot the results of spatial density distribution from numerical analysis for system size $L=50$. It can be noticed that the  system initially has an extended nature of the condensate at a weak $\tilde{V}$ value. The system exhibits enhanced localization with increasing tilt. This is marked by a gradual enhancement of the peak of the density  and the disappearance of the condensate density at the right edge.  We plot QFI, $F_Q$, as a function of $\tilde{V}$ for various $L$ in Fig.~\ref{fig:app_all}(e). The obtained scaling exponents $(\beta, \nu, \tilde{V}_c)\sim(4.7,0.42, 2\times 10^{-4})$ are nearly identical to the case of V-shaped lattice.

%\nocite{*}

\bibliography{apssamp}% Produces the bibliography via BibTeX.

\end{document}